\documentclass[12pt]{article}
\input epsf
\begin{document}
\thispagestyle{empty}

\centerline{\Large\bf  Effect of Hadron Dynamics on the Proton Lifetime}
\bigskip
\centerline{\large Alfred Scharff
Goldhaber*\footnote{goldhab@insti.physics.sunysb.edu}, T.
Goldman$^\dagger$\footnote{t.goldman@post.harvard.edu}, and Richard R.
Silbar$^\dagger$\footnote{silbar@lanl.gov}} 

\bigskip

\centerline{\it *C.N. Yang Institute for Theoretical Physics, }
\centerline{\it State University of New York, Stony Brook, NY 11794-3840}

\smallskip

\centerline{\it $^\dagger$Theoretical Division, Los Alamos National Laboratory,}
\centerline{\it Los Alamos, NM 87545}

\begin{abstract}

A detailed, quantitative re-examination of the effect of hadron
dynamics on baryon decay, modeled in terms of Skyrme-field
tunneling, indicates that any hadronic suppression should be
quite mild.  This appears to be another illustration of the `Cheshire-cat'
phenomenon, that variation of the apportionment between description of
the nucleon as a bag of quarks and description as a Skyrme field configuration
has little influence on many nucleon properties.  Perhaps the largest remaining
uncertainty in evaluating the decay rate has to do with the overlap
between a 
specified quark-antiquark configuration and a final meson state.

\end{abstract} 

%\today, RRS


\section{Introduction}

There is a limited number of known ways to seek physics beyond the
standard model of electroweak and strong interactions.  Increasing
collision energies in laboratory experiments might reveal new
particles.  There may be small effects or rare processes at low
energies.  There might be particles of such high mass that they could
not be created in the laboratory but have survived from an early,
high-temperature epoch in the universe.  Recently there has been
evidence for new physics in the second category: the strong
experimental indication of neutrino flavor oscillations, which imply
small but nonzero neutrino masses \cite{superk,sno}.  This evidence
encourages further pursuit of other phenomena, including proton decay,
which like neutrino mass is not forbidden by some established principle
such as gauge invariance.

During a period of (unsuccessful) search for baryon number
non-conser\-va\-tion in large underground detectors, the question arose
whether the rearrangement of hadronic degrees of freedom required for
decay might substantially inhibit the process.  It was suggested that
such `hadronic quenching' might account for several orders of magnitude
suppression of baryon decay \cite{GGN}.  Recent theoretical studies
guided by the neutrino masses and by the assumption of `grand
unification' of electroweak and strong interactions at some high scale
have led to new predictions of baryon decay, involving different
dominant channels from those first discussed, and smaller coefficients
for the effective four-fermion operators generating the decay
\cite{BPatiWilczek}.  Evidently, if there were several orders of
magnitude further suppression due to hadron dynamics, the prospects for
experimental observation would be dim.  Accordingly, it seems timely to
make a more systematic and quantitative approach to the issue of
hadronic suppression, and we attempt that here.

In principle one would like to carry out a path-integral computation
over the QCD degrees of freedom, starting with a nucleon and ending
with a meson and lepton. However, this remains a daunting task, because
there is such poor quantitative control on the properties of QCD in
exactly this regime, phenomena involving energy scales of the same
order as $\Lambda_{\rm QCD}\approx 0.2 \ {\rm GeV}$.  Therefore, as in
the earlier work, we use instead Skyrme's Ansatz \cite{skyrme} for the
nucleon as a topological soliton of the chiral field, and to provide
more microscopic realism allow a bag to occupy the interior region of
the skyrmion, with three valence quarks inside it.  This is the hybrid
chiral bag model \cite{BR}.  To describe hadronic tunneling in the
initial $B=1$ and final $B=0$ states, we take a single collective
variable, the value of the chiral angle at the bag boundary radius $R$,
where, for the skyrmion in equilibrium, the angle would be $\pi/2$,
halfway between the values $0$ at spatial infinity and $\pi$ at the
origin. We estimate the tunneling action by using the Skyrme Ansatz
throughout space, constraining the Ansatz by imposing varying values
for the chiral angle at $R$.

Assuming that the formulation is exactly correct in every respect
except that we are restricting the tunneling to only the single
variable of the chiral angle at the fixed boundary of the bag, the
result of this calculation should be a lower bound on the hadronic
matrix element of the four-fermion operator which reduces baryon and
lepton numbers each by one unit.  It is a lower bound because taking
account of more degrees of freedom should only increase the tunneling
probability, barring cancellations due to opposing phases.

In our earlier work \cite{GGN} it was suggested from crude dimensional
considerations that the tunneling between a final, trivial vacuum state
and a configuration of the chiral field with topological number zero
but mass equal to the nucleon mass could be equivalent to the motion of
a particle with GeV mass through a GeV potential barrier of thickness
about 1 fm.  This immediately gives a suppression factor $10^{-2}$ in
amplitude or $10^{-4}$ in rate.  Tunneling in the initial $B=1$ state
was ignored.  What we find now is that both the initial and final state
tunneling factors are relatively large, meaning that the Skyrme Ansatz
indicates a soft degree of freedom associated with change of the chiral
angle at the bag radius, either for fluctuations about the vacuum as in
the final state (which consists of vacuum plus a departing meson and
lepton), or for fluctuations about the $B = 1$ initial
state.

Put differently, quantum fluctuations of the vacuum are easily
generated in which objects appear which consist of a quarter skyrmion
outside the radius $R$ and a quarter anti-skyrmion inside (or vice
versa).  In addition, fluctuations of a skyrmion are easily generated
in which one quarter of the baryon number shifts inside (or outside)
the radius $R$. These are not the only modifications from a traditional
(MIT bag \cite{MIT}) estimate of baryon decay.  Because the radius of
the chiral bag boundary is about half that of the MIT bag, one must pay
attention to dimensional dependence of the four-fermion-operator matrix
element on the volume which bag wave functions of quarks occupy.  Also,
the wave functions for the valence quarks take a different form as the
chiral boundary angle varies from the case for chiral boundary angle
$0$, as in the MIT bag.

All these effects are taken into account in what follows.  We begin by
defining the matrix element to be estimated, then determine the initial
and final state tunneling factors, and end by putting in factors for
the initial quark and final meson and lepton wave functions.  It will
be seen that the calculation is nearly identical for the old dominant
channel, $p\to e^+\pi^0$, and for the channel favored in more recent
calculations $p\to \bar{\nu}K^+$, except for the larger mass and hence
lower final velocity of $K$ compared to $\pi$, and the gamma matrices
appearing in the four-fermion operator, neither of which alters the
hadronic tunneling.  Our conclusion is that, because of the large
spontaneous chiral fluctuations, hadronic tunneling implies at most
a quite minor suppression of baryon decay.
\eject

\section{The Decay Matrix Element}

The proton decay amplitude for $p\rightarrow m \bar{\ell}$ in terms of
initial and final state wave functions $\Psi_{\rm final}$ and $\Psi_{\rm
initial}$ is
\begin{eqnarray} 
        <m\bar{\ell}|H_{I}|p> & = 
& \int ds\, \int d^4 x \, 
                \Psi^*_{\rm final}(s,x) \Psi_{\rm initial}(s,x)\, 
\nonumber \\
        & & \quad\times <q\bar{q}\bar{\ell}|\psi^4(x)|q^3>_s \nonumber \\ 
        & \approx & <q\bar{q}\bar{\ell}|\psi^4(x)|q^3>_{s=1, x=0} {\cal I} \ ,
        \label{eq:decayME}
\end{eqnarray} 
where ``initial'' means a Skyrmion with topological number one and ``final''
means vacuum with topological number zero.  Further,
\begin{equation}
         \psi^4 = \sum_i \bar\psi_l {\cal O}_{ia} \psi_{qa}
                        \bar\psi_{qd}^c  {\cal O}_{db}^i \psi_{qb} \Gamma^{abd}
        \label{eq:psi4} \ \ ,
\end{equation} 
and
\begin{equation}
         {\cal I} = \int ds\, 
                \Psi^*_{\rm final}(s) \Psi_{\rm initial}(s)
        \label{eq:Idef}
\end{equation} 
is the inhibition factor due to hadron dynamics that will be calculated
in the next section.  The integration variable $s$  accounts for the
quantum fluctuations (to be defined below in terms of separate scaling
variables, $s_{\rm int}$ and $s_{\rm ext}$, for the initial and final
state wave functions, respectively). The $\Psi$'s in Eq.(\ref{eq:Idef})
do not show an $x$-dependence because that has already been integrated
out in the Skyrme model we use, also to be discussed below.  However,
${\cal I}$ does include a phase factor $e^{-iMt}$, not shown, because
the initial chiral field wave function has the mass of the nucleon,
$M$, while the final (vacuum) wave function has exactly zero energy.
This factor assures that the total energy of the final lepton and meson
will be the mass of the initial nucleon. The approximate factorization
of the amplitude indicated in the last line of Eq.(\ref{eq:decayME})
will be justified in Sec.\ 4.

\section{Calculation of Inhibition Factor from\\ 
        Skyrmion Dynamics}

\subsection{Skyrmion Profile Function}

The Skyrme model of the nucleon \cite{skyrme} begins with a Lagrangian
\begin{equation}
        {\cal{L}} = 
        \frac{F_\pi^2}{16} \mbox{Tr} \{ \partial_\mu U \partial^\mu
U^\dagger \} + 
        \frac{1}{32e^2} \mbox{Tr} \{ [L_\mu, L_\nu]^2 \} \ ,
        \label{eq:lagrangian}
\end{equation} where $L_\mu = U^\dagger \partial_\mu U$ and the static
Skyrmion wave
function $U$ is defined as
\begin{equation}
        U = e^{i \vec{\tau} \cdot \hat{\bf\rm r} F(r)} \  .
\end{equation} 
Here $F_\pi$ is the pion decay constant, which is fitted at 129 MeV by
Adkins, Nappi, and Witten \cite{ANW}, smaller than the experimental
value of 188 MeV (in the conventions of ANW). The quartic ``Skyrme
term'' stabilizes the soliton against collapse; its coefficient has a
dimensionless constant $e$ which is fitted at the value 5.45 in ANW.

Extremizing $\cal L$ to get an ``equation of motion'' for the profile
function $F(r)$ yields the non-linear differential equation
\begin{equation}
        \left(\frac{1}{4}r^2 + 2 \sin^2 F\right) F^{\prime\prime} +
\frac{1}{2} r F^\prime
        + \sin 2F \left[(F^\prime)^2 - \frac{1}{4}\right] - \frac{\sin^2
F}{r^2} \sin 2 F
= 0 \ .
        \label{eq:Feqn}
\end{equation} 
The radial coordinate $r$ here is dimensionless, with the corresponding
dimensional coordinate  given by $r/(e F_\pi)$.

For a single nucleon (i.e., baryon number $B=1$) Eq.(\ref{eq:Feqn})
has  boundary conditions  $F(0) = \pi$ and $F(r) \rightarrow a/r^2$ as
$r \rightarrow \infty$.  It turns out that $F'(0)=-1.03812$, falling
off linearly from the origin, and the asymptotic constant $a =
8.63385$.  We shall use variations of $F(r)$ to describe the pion field
in the initial state of the decaying nucleon.

Note that if $F(r)$ is a solution of this equation, then so is $\pi -
F(r)$.   Let $R$ be the point where $F(R )= \pi - F(R) = \pi/2$.  (From
the solution for $F(r)$, $R=1.76$; for ANW's choice of $F_\pi$ and $e$
this corresponds to a distance from the origin of 0.49 fm.) Thus we
also can create a $B=0$ configuration which solves Eq.(\ref{eq:Feqn})
by taking the piecewise combination
\begin{equation}
        \tilde{F}(r) = \left\{ \begin{array}{ll}
        \pi - F(r) & \mbox{if $r < R$} \ ,\\
        F(r) &  \mbox{otherwise} \ .
        \end{array}
        \right. 
        \label{eq:FtildeDef}
\end{equation} 
This solution is valid except at $r=R$, where the derivative of $F$ is
discontinuous.  A smoother approximate $B=0$ solution with the
constraint $F(R)=\pi/2$ could be obtained easily using
Eq.(\ref{eq:FtildeDef}) by some rounding in the neighborhood of the
cusp at $r=R$.  We shall be using $\tilde{F}(r)$ for the contribution
to the inhibition factor $\cal I$ from  the $B=0$ final state wave
function's chiral field dependence.

The nucleon mass $M$ in the Skyrmion picture \cite{ANW} is just the
value of ${\cal L}$ evaluated for the solution $F(r)$,
\begin{eqnarray}
        M & = & M_2 + M_4       \label{eq:Nmass} \\
        M_2 & = & \frac{\pi F_\pi^2}{2}\left(\frac{1}{eF_\pi}\right)
\int_0^\infty dr\, 
                \left[r^2 F'^2 + 2 \sin^2 F\right] \nonumber \\
        M_4 & = & \frac{2\pi}{e^2} \left(eF_\pi\right)\int_0^\infty dr\,
\sin^2 F
                \left[\frac{\sin^2 F}{r^2} + 2F'^2\right]       
                \nonumber \ .
\end{eqnarray} 
Because we minimized $\cal L$ to get $F$, $M_2 = M_4$.  Note that both
$F(r)$ and $\tilde{F}(r)$ give the same value for $M$. This means that
the $B=0$ state represented by $\tilde F$ does not have an energy
appropriate to the hadronic part of a final state of, say,  a pion and
an electron, much less to the energy of the vacuum. This assumes one
wishes to describe the final meson as produced from a quark-antiquark
pair left by the action of the four-fermion operator on the initial
baryon state, rather than by evolution of the classical chiral field
represented by $\tilde F$.

\subsection{Varying the Initial and Final State Wave Functions}

The Skyrmion is a classical soliton solution which minimizes the
Lagrangian, Eq.(\ref{eq:lagrangian}).   We want to allow for quantum
fluctuations in the calculation of the ``tunneling''  inhibition factor
$\cal I$, and we do that by making variations of $F(r)$ and $\tilde
F(r)$.  We need to do this differently for the two cases, because of
the different boundary conditions to be satisfied, $B=1$ and $B=0$,
respectively. We introduce scaling factors $s_{\rm int}$ and $s_{\rm
ext}$, respectively, which are allowed to depend on time.  The time
derivatives of each $s(t)$ in the modified Lagrangian,
Eq.(\ref{eq:lagrangian}), then lead to Schr{\"o}dinger-like equations
with potential wells that can be reasonably well approximated by
quadratic functions in each $s$. This then lets us write the initial
and final states $\Psi^*_{\rm final}(s_{\rm ext})$ and $\Psi_{\rm
initial}(s_{\rm int})$ as harmonic oscillator wave functions.  In the
end, we can evaluate a simple integral over $s_{\rm ext}$ to get the
inhibition factor $\cal I$.

\subsection{The Final State Wave Function}

\begin{figure}[tb]
        \centering
        \epsffile{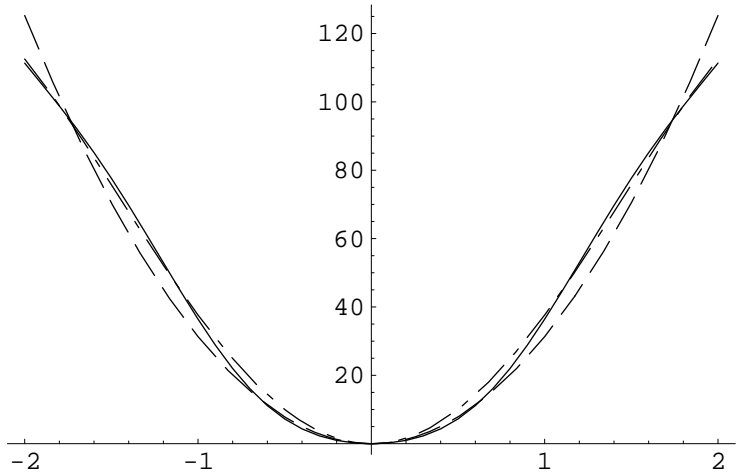}
        \caption{Plot of $V_{\rm ext}(s_{\rm ext})$ in units of $F_\pi/e$
	(solid curve).  The dot-dashed curve is a quartic fit to 
	$V_{\rm ext}(s_{\rm ext})$ and the long-dashed curve is a
	quadratic fit.}
        \label{Vsextfig}
\end{figure}

For purposes of presentation, it is easier to discuss first the scaling
of the final state $\tilde F$. The $B=0$ boundary condition, that
$\tilde F$ vanish at both $r=0$ and $r=\infty$, can be simply
maintained by letting
\begin{equation}
        \tilde F(r) \rightarrow s_{\rm ext}\tilde F(r) \ ,
        \label{eq:extnlscaling}
\end{equation} 
where the subscript ``ext'' refers to scaling external to the
functional argument.  Substituting this in the Lagrangian of
Eq.(\ref{eq:lagrangian}) then gives a ``potential'' $V_{\rm ext}(s_{\rm
ext})$ with a minimum at $s_{\rm ext}=0$, as shown in
Fig.~\ref{Vsextfig}.
The figure shows the calculated $V_{\rm ext}$, which exhibits slight
turnovers at $s=\pm 1.5$, together with a quartic fit to it (hardly
distinguishable) and a quadratic fit (lacking the turnovers).   As
eventually we shall be interested in values of $s_{\rm ext}$ between 0
and 1, the quadratic fit is adequate for our purposes.  The fitted
quadratic is
\begin{equation}
        V_{\rm ext}(s_{\rm ext}) = \frac{1}{2} a s_{\rm ext}^2, \quad a
= 63  \ ,
\end{equation} 
with the factor of $\frac{1}{2}$ inserted to mimic the simple harmonic
oscillator potential.  Here $a$ has units $F_\pi/e$.

\begin{figure}[tb]
        \centering
        %\vspace{2 in}
        \epsffile{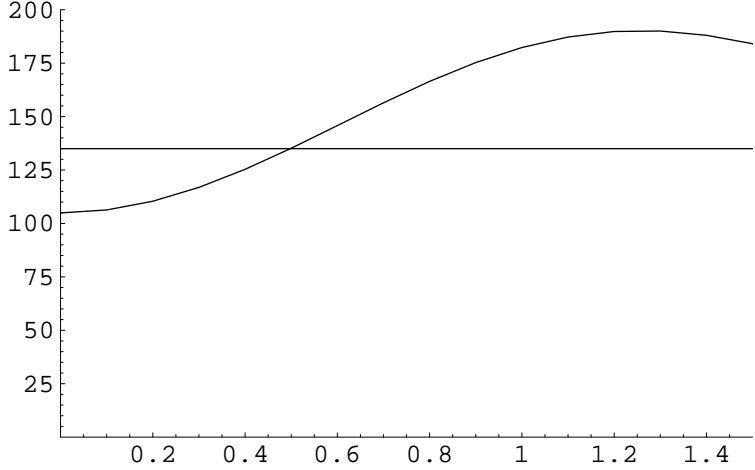}
        \caption{Plot of $I_{\rm ext}(s_{\rm ext})$ in units of $1/e^3F_\pi$.
           The horizontal line has a $y$-axis value of 135, which we 
           will take as a constant approximation for $I_{\rm ext}$.}
        \label{Isextfig}
\end{figure}

If we now go on to allow $s_{\rm ext}$ to have a time dependence, the
Lagrangian in Eq.(\ref{eq:lagrangian}) leads, in both terms (those
quadratic and quartic in derivatives), to a dependence on $\dot{s}_{\rm
ext}^2$,
\begin{eqnarray}
    {\cal L} \rightarrow {\cal L}(s_{\rm ext}) &=& \frac{1}{2}
    I_{\rm ext}\dot{s}_{\rm ext}^2
            + V_{\rm ext}(s_{\rm ext}) \label{eq:Lsextoft} \\
    I_{\rm ext} &=& \frac{\pi}{e^3F_\pi}\int_0^\infty dr\,  r^2
    \tilde F^2(r)\, 
    \left[1 + 8\frac{\sin^2(s_{\rm ext} \tilde F(r))}{r^2}\right]
\nonumber \ .
\end{eqnarray} 
The $s_{\rm ext}$-dependence of the ``moment of inertia'' $I_{\rm ext}$
is shown in Fig.~\ref{Isextfig}.
The variation of $I_{\rm ext}$ over the range $s_{\rm ext} = 0$ to $1$
is small,  so for simplicity we take it to be a constant, $I_{\rm ext}
= 135$ in units of $1/e^3F_\pi$.

The ``momentum'' canonical to $s_{\rm ext}$ is $(I_{\rm ext})
\dot{s}_{\rm ext}$.  Converting $\cal L$ into a Hamiltonian and
replacing the momentum by $-i \partial / \partial s_{\rm ext}$ then
gives a Schr{\"o}\-dinger-like equation for the final state (vacuum)
wave function,
\begin{equation}
  \left[- \frac{1}{2I_{\rm ext}} \frac{\partial^2}{\partial s_{\rm
ext}^2} + 
    \frac{1}{2} a s_{\rm ext}^2\right]\Psi_{\rm ext}(s_{\rm ext})  
                = E \Psi_{\rm ext}(s_{\rm ext}) \ .
        \label{eq:extSEqn}
\end{equation} 
This equation has the form of the familiar harmonic oscillator
problem.  We take $\Psi_{\rm final}$ to be the ground state
eigenfunction,
\begin{equation}
        \Psi_{\rm final}(s_{\rm ext}) = N_{\rm ext} 
        \exp(-\alpha_{\rm ext}^2 s_{\rm ext}^2/2)  \ .
        \label{eq:Psi_vac}
\end{equation} 
where $\alpha_{\rm ext}^4 = ``m K$'' $= I a/e^4$.   For ANW's $e =
5.45$, we get $\alpha_{\rm ext}^2 = 3.105$ and the normalization
constant $N_{\rm ext}^2 = \alpha_{\rm ext}/\sqrt\pi = 0.994$.   The
ground state eigenvalue is
\begin{equation}
        E_0 = \frac{1}{2} \sqrt(a/I) = 0.342
        \label{eq:EZEROPsi_vac}
\end{equation} 
in units of $F_\pi e$, i.e $\approx  240$ MeV, using ANW's fitted
values for these coupling constants.

%This is not far from the energy
%we might expect for the final-state hadron.

\subsection{The Initial State Wave Function}

For the initial state, in making variations of the $F$ we need to
maintain the $B=1$ boundary conditions mentioned below
Eq.(\ref{eq:Feqn}).   We therefore have chosen simply to introduce
another scaling parameter $s_{\rm int}$ so that
\begin{equation}
        F(r) \rightarrow F(s_{\rm int}r) \ ,
        \label{eq:intnlscaling}
\end{equation} 
where the subscript ``int'' refers to scaling internal to the function
$F(r)$. It will turn out that the scaling variable here, $s_{\rm int}$,
is not quite what is needed for the integration over $s$ in
Eq.(\ref{eq:decayME}), which we choose in Eq.(\ref{eq:Idef}) to be the
$s_{\rm ext}$ defined in the last sub-section. The relation between the
two scaling variables will be dealt with below.

\begin{figure}[tb]
        \centering
        \epsffile{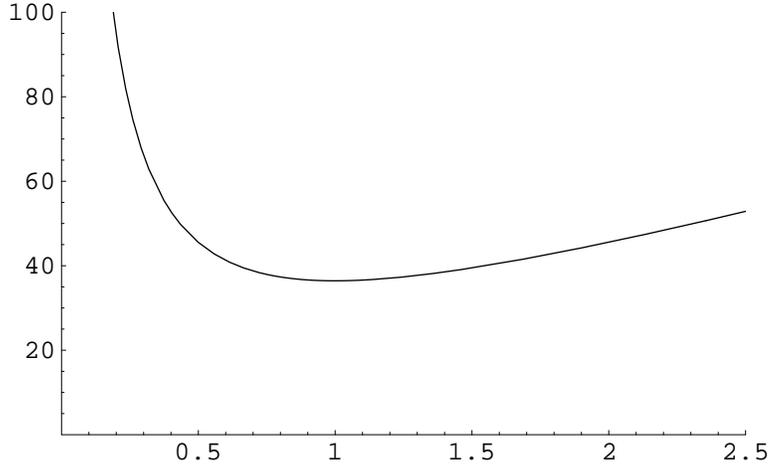}
        \caption{Plot of $V_{\rm int}(s_{\rm int})$ around $s_{\rm int}=1$
        in units of $F_\pi/e$.}\label{Vsintfig}
\end{figure}

Using $F(s_{\rm int}r)$ we can  define a ``potential'' for the initial
state wave function from the expression for the nucleon mass,
Eq.(\ref{eq:Nmass}):
\begin{equation}
        V_{\rm int}(s_{\rm int}) = M[F(s_{\rm int}r)] = 
        M_2/s_{\rm int} + M_4 s_{\rm int}
        \label{eq:intVdef}
\end{equation} 
This simple $s_{\rm int}$-dependence can be understood by replacing
$r$, the variable of integration in Eq.(\ref{eq:Nmass}), by $r' =
s_{\rm int}r$.  Figure~\ref{Vsintfig} shows that $V(s_{\rm int})$ has a
shallow minimum at $s_{\rm int}=1$, as it should.

Again, we let $s_{\rm int}$ have a time dependence and find the
modified Lagrangian to be
\begin{eqnarray}
        {\cal L} \rightarrow {\cal L}(s_{\rm int}) &=& 
                \frac{1}{2}I_{\rm int}(s_{\rm int})\dot{s}_{\rm int}^2
                + V_{\rm int}(s_{\rm int}) \label{eq:Lsintoft} \ ,\\
        I_{\rm int}(s_{\rm int}) &=& I_{\rm int, 2}(s_{\rm int}) + 
                I_{\rm int, 4}(s_{\rm int})  \label{eq:Isint} \ ,  \\
        I_{\rm int, 2}(s_{\rm int}) &=&
        \frac{\pi}{e^3F_\pi}\int_0^\infty dr\, r^4 
                \tilde F'^2(s_{\rm int}r) \ , \\
        I_{\rm int, 4}(s_{\rm int}) &=&
        \frac{4\pi}{e^3F_\pi}\int_0^\infty dr\,  r^2 
                \tilde F'^2(s_{\rm int}r)\, \sin^2(\tilde F(s_{\rm
        int}r))  \\
                & & \quad\quad\quad\times\left[ 1 + 2\sin^2(\tilde
        F(s_{\rm int}r)) \right] \nonumber \ .
\end{eqnarray} 
Note that $I_{\rm int, 2}$'s integrand has a slow fall-off in $r$,
which necessitates some care in calculating the contribution of the
asymptotic tail. By the same change of integration variable as for
$V_{\rm int}$ we see that
\begin{equation}
        I_{\rm int}(s_{\rm int}) = I_{\rm int, 2}(1)/s_{\rm
        int}^5 + I_{\rm int, 4}(1)/s_{\rm int}^3 \ . 
        \label{eq:Isintdepnce}
\end{equation} 
Evaluating the integrals in Eq.(\ref{eq:Isint}), we find $I_{\rm int,
2}(1) = 110.46$ and $I_{\rm int, 4}(1) = 46.86$.

One might worry that $I_{\rm int}$ is a steeply falling function of
$s_{\rm int}$.  However, as already mentioned, we shall be doing the
final integration in Eq.(\ref{eq:Idef}) over $s_{\rm ext}$. Thus we
need the relation between the two scaling variables, which is
\begin{eqnarray}
        s_{\rm ext} & = & \frac{2}{\pi} F(s_{\rm int}R) \ . \label{eq:srelation}
\end{eqnarray} 
This relation also defines, implicitly, $s_{\rm int}$ as a function of
$s_{\rm ext}$. Note that as $s_{\rm int}$ runs from 0 to 1 to $\infty$,
$s_{\rm ext}$ falls from 2 to 1 to 0. From Eq.(\ref{eq:srelation}) we
also have
\begin{equation}
        \frac{ds_{\rm ext}}{dt} = \frac{2R}{\pi} 
        \left[\frac{dF(r)}{dr}\right]_{r=s_{\rm int}R}
                 \frac{ds_{\rm int}}{dt} \ .    \label{eq:dsrelation}
\end{equation}

\begin{figure}[tb]
        \centering
        %\vspace{1 in}
        \epsffile{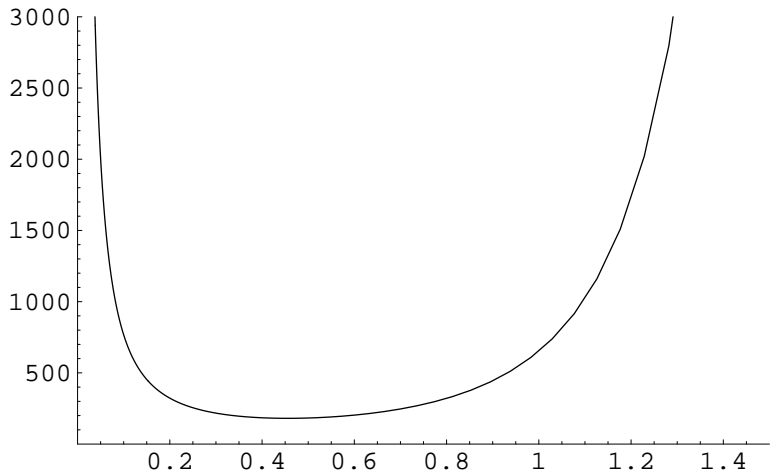}
        \caption{Plot of $X$ as a function of $s_{\rm ext}$
        in units of $1/e^3F_\pi$.}\label{Isintvssext}
\end{figure}

We need this last relation, Eq.(\ref{eq:dsrelation}), for converting
the ``kinetic energy''  term in Eq.(\ref{eq:Lsintoft}). With it, the
moment of inertia term as a function of $s_{\rm ext}$ becomes
\begin{eqnarray}
        \frac{1}{2}I_{\rm int}(s_{\rm int})\dot{s}_{\rm int}^2 & = & 
                \frac{1}{2} X(s_{\rm int}) \left( \frac{ds_{\rm
        ext}}{dt}\right)^2 \ , \\
                X(s_{\rm ext}) & = &
                \left( \frac{\pi}{2R} \right)^2 \left[\frac{1}{dF(s_{\rm
        int}R)/dr}\right]^2
                I_{\rm int}(s_{\rm int}) \ ,    \label{eq:momintterm}
\end{eqnarray} 
where Eq.(\ref{eq:srelation}) is used to convert the variable $s_{\rm
int}$ to $s_{\rm ext}$. Since $F'(s_{\rm int}R)$ falls off like
$1/s_{\rm int}^3$ at large $s_{\rm int}$,  the $1/F'\,^2$ factor more
than compensates the $1/s_{\rm int}^3$ falloff of $I_{\rm int}(s_{\rm
int})$ seen in Eq.(\ref{eq:Isintdepnce}). Thus $X(s_{\rm ext})$ is a
steep-sided, somewhat flat-bottomed well  (shown in
Fig.~\ref{Isintvssext}) instead of a steeply falling function.
Below we shall  replace $X(s_{\rm ext})$ by a constant, $X_0$, which we
take as its value over the flat bottom.  $X_0 \approx 300$ in units of
$1/e^3F_\pi$. This is a reasonable approximation in that the wave
function we are trying to find, $\Psi_{\rm initial}$, will be
exponentially small in the regions where $X(s_{\rm ext})$ is large.

\begin{figure}[tb]
        \centering
        %\vspace{1 in}
        \epsffile{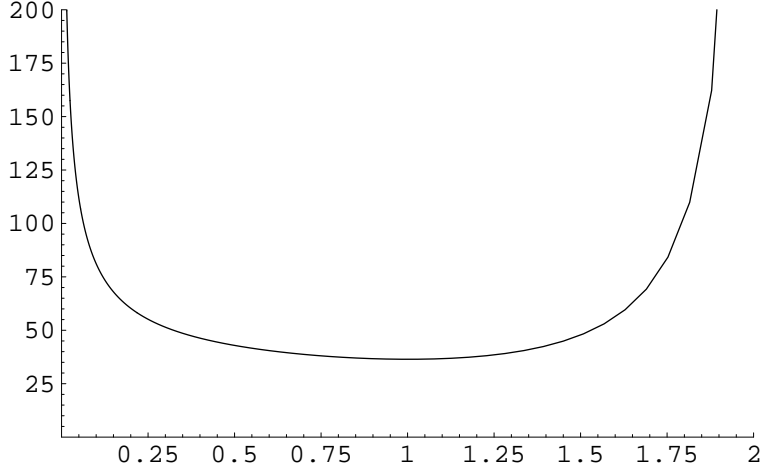}
        \caption{Plot of $V_{\rm int}(s_{\rm int})$ as a function of
        $s_{\rm ext}$
        in units of $F_\pi/e$.}\label{Vintsext}
\end{figure}

In addition to the kinetic energy term, we also need to convert the
potential term into a function of $s_{\rm ext}$.  This also undergoes a
dramatic change of form from that in Fig.~\ref{Vsintfig}, as shown in
Fig.~\ref{Vintsext}.
$V_{\rm int}(s_{\rm int})$ still has is minimum at $s_{\rm ext} =
s_{\rm int} = 1$, as it must.   The steep rise near $s_{\rm ext} = 2$
reflects the $1/s_{\rm int}$ fall-off near $s_{\rm int} = 0$. The rise
near $s_{\rm ext} = 0$ results from $s_{\rm ext} \sim s_{\rm int}^{-2}$
as $s_{\rm int}$ gets large. That is, for small $s_{\rm ext}$, the
linear growth of $V_{\rm int}$ in $s_{\rm int}$ for large $s_{\rm int}$
turns into $s_{\rm ext}^{-1/2}$ behavior near $s_{\rm ext} = 0$.

As in the last section, we can now convert the modified Lagrangian in
Eq.(\ref{eq:Lsintoft}) into a Hamiltonian and thence into a Schr{\"
o}dinger-like equation for the initial-state wave function $\Psi_{\rm
initial}$,
\begin{equation}
        \left[- \frac{1}{2X(s_{\rm ext})} \frac{\partial^2}{\partial
        s_{\rm ext}^2} + 
           V_{\rm int}(s_{\rm ext})\right]\Psi_{\rm int}(s_{\rm ext})  
                = E \Psi_{\rm int}(s_{\rm ext}) \ .
        \label{eq:intSEqn}
\end{equation}

\begin{figure}[tb]
        \centering
        %\vspace{1 in}
        \epsffile{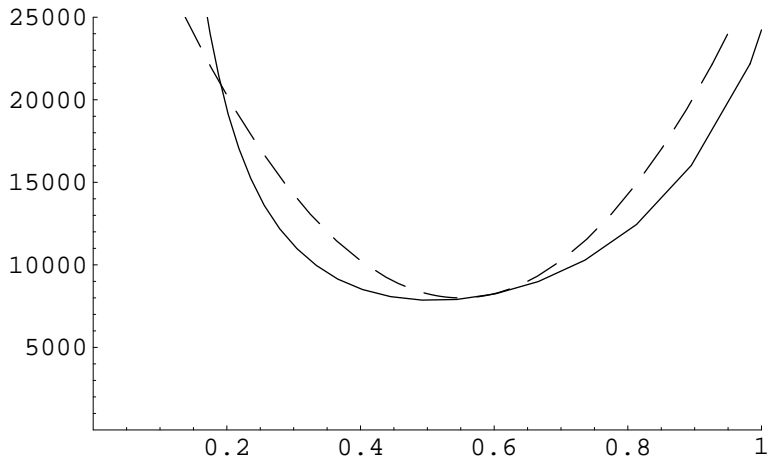}
        \caption{Plot of $XV$ as a function of $s_{\rm ext}$ (solid curve),
                compared with the quadratic fit (dashed curve) described 
                in the text.}
        \label{XVsintvssextfit}
\end{figure}

We can recast Eq.(\ref{eq:intSEqn}) into an (approximate) harmonic
oscillator equation on multiplying up the $X(s_{\rm ext})$ and
approximating $X(s_{\rm ext}) E \Psi_{\rm int}(s_{\rm ext})$ by
$X_0E\Psi_{\rm int}(s_{\rm ext})$. The product of $X(s_{\rm ext})V_{\rm
int}(s_{\rm ext})$ is shown as the solid curve in
Fig.~\ref{XVsintvssextfit}.  It looks a little lopsided, but we can
find an approximate quadratic fit to $XV$ as
\begin{equation}
        X(s_{\rm ext})V_{\rm int}(s_{\rm ext}) \approx 
                [b + \frac{1}{2}c(s_{\rm ext} - {\bar s})^2]
        \label{eq:XVfitn}
\end{equation} 
with $b=8,000$, $c= 200,000$ in units of $1/e^4$ and ${\bar s} = 0.55.$
The quality of this fit is shown in Fig.~\ref{XVsintvssextfit}.
The constant term $b$ just represents a shift in energy, so we want the
solution to
\begin{equation}
        \left[- \frac{1}{2} \frac{\partial^2}{\partial s_{\rm ext}^2} + 
                \frac{1}{2}c(s_{\rm ext} - {\bar s})^2 \right]\Psi_{\rm
        int}(s_{\rm ext})  
                = X_0 E \Psi_{\rm int}(s_{\rm ext}) \ .
        \label{eq:intSEqn2}
\end{equation} 
As before, we take $\Psi_{\rm initial}$ to be the ground state
eigenfunction,
\begin{equation}
        \Psi_{\rm initial}(s_{\rm ext}) = N_{\rm int} 
         \exp(-\alpha_{\rm int}^2 (s_{\rm ext} - {\bar s})^2/2) \ .
        \label{eq:Psi_initial}
\end{equation} 
where $\alpha_{\rm int}^4 = ``m K$'' $= c/e^4$.   For ANW's $e = 5.45$,
we get $\alpha_{\rm int}^2 = 15.06$ and the normalization constant
$N_{\rm int}^2 = \alpha_{\rm int}/\sqrt\pi = 2.189$.

\subsection{Calculating the Overlap Integral}

With the wave functions given in Eq.(\ref{eq:Psi_vac}) and
(\ref{eq:Psi_initial}), the integral over $s = s_{\rm ext}$ in the
definition of the inhibition factor $\cal I$, Eq.(3), can be readily
evaluated to give
\begin{equation}
        {\cal I} = \sqrt{\frac{2\alpha_{\rm ext}\alpha_{\rm int}}
                {\alpha_{\rm ext}^2+\alpha_{\rm int}^2}}
                \exp[-\frac{\alpha_{\rm ext}^2\alpha_{\rm int}^2}
                {2(\alpha_{\rm ext}^2+\alpha_{\rm int}^2)}{\bar s}^2] \ .
        \label{eq:Ievaluated}
\end{equation} 
With the values of $\alpha_{\rm ext}$,  $\alpha_{\rm int}$, and $\bar
s$ obtained above for the ANW choices of $F_\pi$ and $e$, this
evaluates to
\begin{equation}
        {\cal I} = 0.588 \ .
        \label{eq:Inumber}
\end{equation}

\subsection{The Initial State Wave Function, Reconsidered}

When we first saw the result for $\cal I$ given in the last section, it
was a major surprise.  We had expected the value to be quite a bit
smaller, even if not so small as indicated in Ref.\cite{GGN}.  Thus, we
felt compelled to calculate the initial state wave function $\Psi_{\rm
initial}$ in a second, more systematic way.

\begin{figure}[tb]
        \centering
        %\vspace{1 in}
        \epsffile{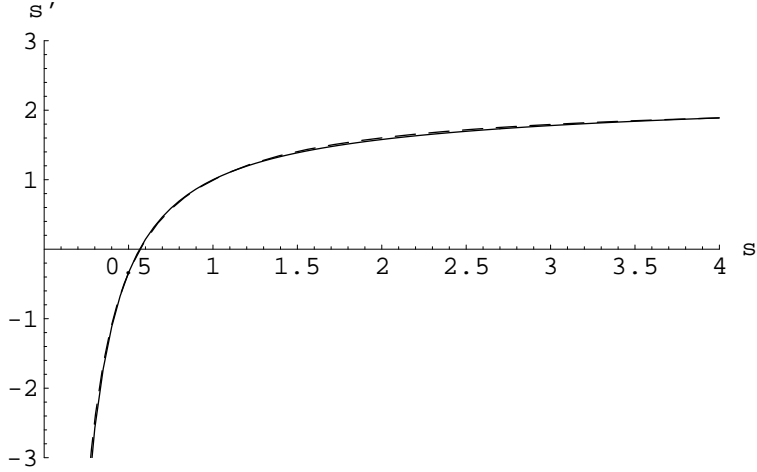}
        \caption{Plot of $s'$ as a function of $s=s_{\rm int}$,
                compared with its hyperbolic fit (dashed curve).}
        \label{spr_of_s}
\end{figure}

Let us convert the first term of Eq.(\ref{eq:Lsintoft}) to something
more tractable by defining a new variable $s'$ so that
\begin{equation}
        I_2 \left(\frac{d s'}{d t}\right)^2 = 
        I_{\rm int}\left(\frac{d s_{\rm int}}{d t}\right)^2 \equiv 
        I_2 \  g(s_{\rm int}) \left(\frac{d s_{\rm int}}{d t}\right)^2,
        \label{eq:sprdef}
\end{equation}
where
\begin{equation}
        g(s) = [1 + \lambda s^2]/s^5 \ , \quad 
        \lambda = I_{{\rm int},4}/I_{{\rm int},2} = 0.4242 .
        \label{eq:gdef}
\end{equation}
As a function of $s_{\rm int}$, $s'$ is then given by 
\begin{equation}
        s'(s_{\rm int}) = \int_1^{s_{\rm int}} \sqrt{g(s)}\ ds ,
        \label{eq:sprfcn}
\end{equation}
with $s' = 1$ when $s_{\rm int} = 1$.  A plot of $s'(s_{\rm int})$ is
shown in Fig.~\ref{spr_of_s}.
Note that $s'$ asymptotes to 2.55.  It can be fit well with a
hyperbola,
\begin{equation}
        s'(s_{\rm int}) = a + b/(s_{\rm int} - c) ,
        \label{eq:sprfcnfit}
\end{equation}
where $a = 2.167$, $b = -1.093$, and $c = 0.06351$.  This fit is easily
inverted to find $s_{\rm int}$ as a function of $s'$,
\begin{equation}
        s_{\rm int}(s') = c + b/(s' - a) .
        \label{eq:sintfcnfit}
\end{equation}
The fits in Eqs.(\ref{eq:sprfcnfit}) and (\ref{eq:sintfcnfit}) are,
because of the singularities at $s' = a$ and $s_{\rm int} = c$, only
valid over a limited range between the singularities.  This is not a
serious problem as we are only interested in variations of $F(r)$
around $s' = s_{\rm int} = 1$.  As $s'$ runs, say, from 0 to 2, $s_{\rm
int}$ varies from 0.5678 to 6.612, steeply rising near $s' = 2$.

\begin{figure}[tb]
        \centering
        %\vspace{1 in}
        \epsffile{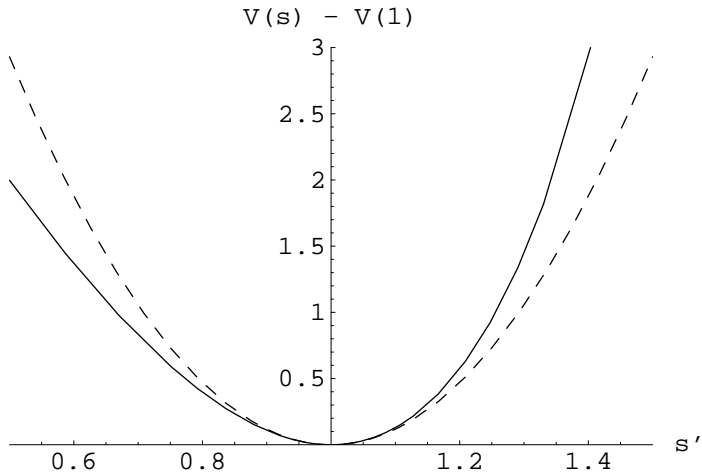}
        \caption{Plot of $V[s_{\rm int}(s')] - V[s'=1]$ around $s' = 1$.
        Units are in $F_\pi/e$. The dashed curve is a quadratic fit.}
        \label{V_of_spr}
\end{figure}

We are now in a position to write a Schr{\"o}dinger-like equation for
$\Psi_{\rm int}$ in terms of the $s'$ variable,
\begin{equation}
        \left\{ -\frac{1}{2I_2} \frac{d^2}{d s'^2} + 
                V_{\rm int}[s_{\rm int}(s')] \right\} \Psi_{\rm int}(s')
                = E \Psi_{\rm int}(s') \ .
        \label{eq:sprSEqn}
\end{equation}
The function $V_{\rm int}[s_{\rm int}(s')]$ has a shallow
non-symmetrical minimum at $s' = 1$, as shown in Fig.~\ref{V_of_spr}.

Let us once again fit $V(s')$ to a quadratic (by evaluating the second
derivative of $1/s_{\rm int}(s') + s_{\rm int}(s')$ with respect to
$s'$), finding
\begin{equation}
        V_{\rm int}[s_{\rm int}(s')] = 
        V(1) + \frac{1}{2} K_{\rm int} (s'-1)^2 \ ,
        \label{eq:VofsprFit}
\end{equation}
where $K_{\rm int} = 2 \times 0.644 M_2 = 23.44$, in units of
$F_\pi/e$, about a third of the ``spring constant'' for the final
state.  This fit is shown as the dashed curve in Fig.\ \ref{V_of_spr}.
We then approximate the ground state solution to Eq.(\ref{eq:sprSEqn})
as
\begin{equation}
        \Psi_{\rm int}(s') = 
        N_{\rm int} e^{-\alpha'^2 (s'-1)^2/2} \ 
        {\rm for\ } 0 \leq s' \leq 2 \ ,
        \label{eq:Psi_init_spr}
\end{equation}
and we take $\Psi_{\rm int}(s')$ equal to 0 outside that range.  Here
$\alpha'^4 = I_2 K_{\rm int}/e^4$ so $\alpha'^2 = 1.721$ for ANW's
value of $e$.  This value is about half that for $\alpha_{\rm ext}$
found in Sec.\ 3.3, meaning that the initial state is ``softer'' than
the final state.  The Gaussian in Eq.(\ref{eq:Psi_init_spr}) has fallen
to 42\% of its peak value at the ``endpoints'' $s' = 0$ and 2.  The
normalization constant $N_{\rm int}$ is determined by integrating the
square of $\Psi_{\rm int}(s')$ over the range where it is non-zero, and
it turns out to be 0.889.  The ground state eigenvalue for this initial
state wave function is
\begin{equation}
        E_0 = \frac{1}{2} \sqrt{K_{\rm int}/I_2} = 0.230,
        \label{eq:EzeroPsi_init}
\end{equation} 
again in units of $F_\pi e$.  Thus, $E_0 \approx 160$ MeV, using ANW's
fitted values for these coupling constants.

\subsection{Calculating the Overlap Integral, Part II}

With the wave functions given in Eq.(\ref{eq:Psi_vac}) and
(\ref{eq:Psi_init_spr}), the integral over $s_{\rm ext}$ in the
definition of the inhibition factor $\cal I$ is a bit trickier to
evaluate.  We will do the integration this time using the $s'$
variable, which requires knowing how $s_{\rm ext}$ varies as a function
of $s'$.

\begin{figure}[tb]
        \centering
        %\vspace{1 in}
        \epsffile{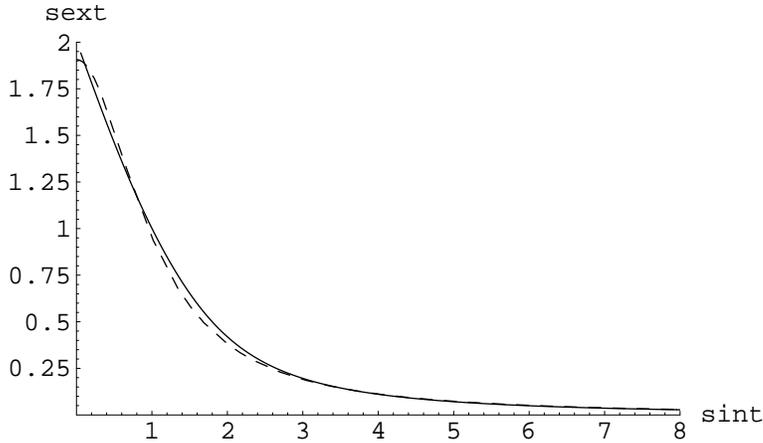}
        \caption{Plot of $s_{\rm ext}$ as a function of $s_{\rm int}$,
                compared with the fit described in the text (dashed curve).}
        \label{sextvssint}
\end{figure}

From Eq.(\ref{eq:srelation}) we show in Fig.~\ref{sextvssint} how
$s_{\rm ext}$ varies as a function of $s_{\rm int}$.
$s_{\rm ext}$ falls from 2 at $s_{\rm int} = 0$ to 0 as $s_{\rm int}^2$
as $s_{\rm int}$ gets large.  Thus a reasonable fit (the dashed curve
in Fig.~\ref{sextvssint}) is given by
\begin{equation}
        s_{\rm ext} = 1.9065/(1 + s_{\rm int}^2) \ .
        \label{eq:sextfit}
\end{equation}

Using the fits in Eqs.(\ref{eq:sprfcnfit}) and (\ref{eq:sextfit}), we
can now evaluate the overlap integral numerically over the range where
$\Psi_{\rm initial}$ is defined,
\begin{equation}
         {\cal I} = \int_0^2 
                \Psi^*_{\rm final}\{s_{\rm ext}[s_{\rm int}(s')]\}
                \Psi_{\rm initial}(s') \, ds' = 0.4975
        \label{eq:IevalNew}
\end{equation} 
We consider this result to be in substantial agreement (and
corroborating) the value for $\cal I$ found in Sec.\ 3.5.

\section{The Hypoweak Matrix Element}

The hypoweak interaction matrix element is the first factor in the
second line of Eq.(\ref{eq:decayME}), the matrix element of the
operator in Eq.(\ref{eq:psi4}).  This has been evaluated for chiral
vector-axial combinations in terms of the wavefunction at the origin,
providing a flux factor for the two quark to lepton plus antiquark
scattering amplitude. As such, the only change needed from previous
work \cite{Ross} is the change in value due to the change in the
standing wave for the quark interior between the MIT bag models used
earlier (with zero chiral angle) and the constant wavefunctions
appropriate to our new chiral boundary condition. The wavefunction
factor becomes
\begin{equation}
        \frac{j_{0}^{2}(0) + j_{1}^{2}(0)}{N^{2}} = 
        \frac{3}{4\pi R^{3}}    \label{eq:piby2}
\end{equation}
where $N$ is the normalization and the spherical Bessel functions of
the MIT bag are replaced by a constant from the fermion upper
wavefunction component and zero from the lower.

Scaling from the result in Ref.\cite{Ross}, where $|\psi(0)|^{2} = 1.1
\times 10^{-3} {\rm GeV}^{2}$ was used, to the value here with $R =
0.6\: {\rm fm}$, we find an enhancement factor of 7.7, which more than
compensates for the overlap suppression from $\cal I$.

For the intermediate case of a boundary condition chiral angle of
$\pi/4$, we need the appropriate frequency for the quark wavefunction
solution.  We need to solve
\begin{equation}
2 j_{0}(\omega R) j_{1}(\omega R) \tan(\theta_{c}) = 
(j_{0}(\omega R))^2 - (j_{1}(\omega R))^2\:,
\end{equation}
which gives the value $\omega R = 1.7446$ for 
\begin{equation}
\theta_{c} = \pi - \theta(R) + \frac{\sin(2\theta(R))}{2} \: .
\end{equation}
Evaluating the usual normalization integral, we find
\begin{eqnarray}
\omega^3 N(\omega)^2 & = & 4 \pi \int_{0}^{\omega R} dx \, x^2
[(j_{0}(x))^2 + (j_{1}(x))^2] 
                 =  14.935 \: .
\end{eqnarray}          
Eq.(\ref{eq:piby2}) then becomes
\begin{equation}
        \frac{j_{0}^{2}(0) + j_{1}^{2}(0)}{N^{2}(\omega)} = 
        \frac{5.3096}{14.935 R^{3}} \: .    \label{eq:piby4}
\end{equation}

This produces an enhancement factor almost three times smaller than
above, but still implies that for the net matrix element including the
tunneling overlap factor ${\cal I}$ there is no major change from the
original MIT bag calculation.

\section{Conclusions}

The fact that hadronic suppression of baryon decay {\it could} be much
less significant than suggested in Ref.\cite{GGN} had become clear to
us some time ago, but we still were surprised to find virtually no
suppression.  It should be emphasized that Ref.\cite{GGN} did not
assert that the suppression would be large, only noting that an
exponential amplitude suppression factor exhibiting a negative exponent
with order of magnitude unity could easily account for two orders of
magnitude reduction in the decay rate.  What has developed in our much
more detailed calculation above is that several effects serve to weaken
this potential suppression.  First, the size of the vacuum
fluctuations, when examined more carefully, turns out to be larger than
simple dimensional counting suggested.  Secondly, the initial nucleon
wave function as well as the final vacuum wave function allows
substantial local fluctuations in baryon density, so that the overlap
between the former (once acted on by the four-fermion operator) and the
latter can be rather large.  Thirdly, the four-fermion matrix element
becomes larger when one considers the initial quarks as confined in a
smaller bag than the MIT bag.

The net effect, within the remaining uncertainties in the calculations,
is that the present more elaborate analysis used here produces
virtually the same prediction as that in the MIT bag model.  This might
be one more example of the ``Cheshire-cat'' phenomenon \cite{chesh,BR},
that many properties of the nucleon are quite insensitive to the choice
of demarcation radius between a chiral field description (used outside
that radius) and a description in terms of quarks (inside), subject to
an appropriate chiral boundary condition. If so, then even the decay of
the proton may not reveal what lies inside it, instead quite directly
measuring the microscopic operator behind the decay.

In this work we have not focused on the matrix element for production
of the final meson from the remaining quark and the produced
antiquark.  Independently of the hybrid bag picture, this might be a
worthwhile subject for additional study, and probably is the largest
source of remaining uncertainty in the lifetime calculation.  Assuming
this also does not produce a large effect, then the latest quoted
limits on both the old $e^+\pi^0$ and new $\bar\nu{\rm K}^+$ favored
channels for proton decay already provide ominous constraints on grand
unified models \cite{jung}.

\section{Acknowledgments}  

Shmuel Nussinov was an active partner in the early stages of this work,
and he was the first to suggest that a more careful look might reveal
much less substantial hadronic suppression of baryon decay than the
old, but crude, arguments had indicated.  Our further efforts have more
than vindicated that view. This research is supported in part by the
Department of Energy under contract W-7405-ENG-36.

\end{document}